\documentstyle[11pt,moriond,epsfig]{article}

\def\be{\begin{equation}}
\def\ee{\end{equation}}
\def\bea{\begin{eqnarray}}
\def\eea{\end{eqnarray}}

\begin{document}
\vspace*{4cm}
\title{BLACK HOLE COLLISIONS: \\
HOW FAR CAN PERTURBATION THEORY GO?}

\author{MANUELA CAMPANELLI}

\address{Max-Planck-Institut fuer Gravitationsphysik, 
Albert-Einstein-Institut, Schlaatzweg 1,\\ D-14473 Potsdam, Germany.}

\maketitle\abstracts{The computation of gravitational radiation 
generated by the coalescence of inspiralling binary black holes 
is nowdays one of the main goals of numerical relativity. 
Perturbation theory has emerged as an ubiquitous tool for all 
those dynamical evolutions where the two black holes 
start close enough to each other, to be treated as single 
distorted black hole (close limit approximation), providing  
at the same time useful benchmarks for full numerical 
simulations. Here we summarize the most recent developments to study 
evolutions of perturbations around rotating (Kerr) black holes. 
The final aim is to generalize the close limit approximation to 
the most general case of two rotating black holes in orbit around 
each other, and thus provide reliable templates for the gravitational 
waveforms in this regime. For this reason it has become very important 
to know if these predictions can actually be trusted to larger 
separation parameters (even in the region where the holes have 
distinct event horizons). The only way to extend the range of 
validity of the linear approximation is to develop the theory of 
second order perturbations around a Kerr hole, by generalizing 
the Teukolsky formalism.}

The prediction of accurate waveforms generated during the final orbital
stage of binary black holes has become a worldwide research topic in general
relativity during this decade. The main reason is that these catastrophic
astrophysical events, considered one of the strongest sources of
gravitational radiation in the universe, are potentially observable by LIGO,
VIRGO and other interferometric detectors, now under construction.
For its strong nonlinear features this black hole merger problem is only
fully tractable by direct numerical integration (with supercomputers) of
Einstein equations. Several difficulties remain to be solved in this
approach such as the presence of early instabilities in the codes for 
numerical evolution of Einstein theory (see E.Seidel's 
contribution to these proceedings for a summary and references on
this problem), and finding a new prescription for astrophysically
realistic initial data representing orbiting black holes.
Meanwhile, perturbation theory has shown not only to be the main approximation
scheme for computation of gravitational radiation, but also a useful
tool to provide benchmarks for full numerical simulations. 
The idea is that one can start a full numerical 
collision evolution with supercomputers and eventually have 
perturbation theory take over in case the numerical evolution 
crashes. From the
theoretical point of view perhaps the more relevant contribution
during the nineties in perturbative theory has been the ``close limit
approximation'' \cite{PP94}. It considers the final merger state of two
black holes as described by a {\it single} perturbed one. This idea was
applied to the head-on collision of two black holes and the emitted
gravitational radiation was computed by means of the techniques used in
first order perturbation theory
around a Schwarzschild black hole. When the results of this computation have
been compared with those of the full numerical integration of Einstein
equations the agreement was so good that it was disturbing \cite{APPSS95}. 
This encouraged the significant effort invested into the development of a 
second order Zerilli formalism of metric perturbations about the Schwarzschild
background. The method was successfully implemented with particular emphasis
on the comparison with the fully numerically generated results. In the case
of two initially stationary black holes (Misner data) the agreement of the
results is striking \cite{GNPP96}. 

%\begin{center}
%\begin{figure*}
%\psfig{figure=fig_4.eps,height=3.3in}
%\caption{As it was shown in Ref. the waveforms are 
%significantly improved by the use of second order perturbation theory. 
%The agreement is excellent for much larger separation than expected, 
%even for this case where the two black holes are really separated 
%$\mu_0=2$ i. e. $L/M\approx 3.3$ being $L$ the distance in the conformal
%space (a common apparent horizon only begins to form at $\mu_0=1.36$).
%\label{fig:waveforms}}
%\end{figure*}
%\end{center}

Second order perturbation theory confirmed
the success of the close limit approximation with an impressive agreement in
both waveforms and energy radiated against the full numerical simulations.
There has been a tantamount success in the extension of these studies to the
case of initially moving towards each other black holes \cite{NGPP98}, and
for slowly rotating ones \cite{GNPP98} (See Ref. \cite{P98} for a
comprehensive review).

All the above close limit computations are based on the Zerilli \cite{Z70}
approach to metric perturbations of a Schwarzschild, i.e. nonrotating, black
hole. This method uses the Regge-Wheeler \cite{RW57} decomposition of the
metric perturbations into multipoles (tensor harmonics). Einstein equations
in the Regge-Wheeler gauge reduce to two single wave equations for the even
and odd parity modes of the gravitational perturbations. There is, however,
the strong belief that binary black holes in a realistic astrophysical
scenario merge together into a single, highly rotating, black hole. There is
also concrete observational evidence of accreting black holes \cite{chinos}
that places the rotation parameter as high as $a/M\simeq $ $0.95$. Finally,
highly rotating black holes provide a new scenario to compare perturbative
theory with full numerical integrations of Einstein equations.

The Regge-Wheeler-Zerilli techniques cannot be extended to study 
perturbations on a Kerr black hole background (see Ref.\ \cite{GNPP98}
for the slowly rotating case). In this case
there is not a multipole decomposition of metric perturbations
(in the time domain) and Einstein equations cannot be uncoupled into wave
equations. A reformulation of the gravitational field equations
due to Newman and Penrose \cite{NP62}, based on the Einstein equations and
Bianchi identities projected along a null complex tetrad, 
$\{l^\mu,n^\mu,m^\mu,\bar{m}^\mu\}$,
allowed Teukolsky\cite{T72} to write down a single master wave
equation for the perturbations of the Kerr metric in terms of the Weyl
scalars $\psi_4$ or $\psi_0$.
This formulation has several advantages: i) It is a first order gauge
invariant description. ii) It does not rely on any multipole decomposition.
iii) The Weyl scalars are objects defined in the full nonlinear theory. 
In addition, the Newman-Penrose formulation
constitutes a simpler and more elegant framework to organize higher
order perturbation schemes as we will see in the next section.

Since the seventies the Teukolsky equation for the first order perturbations
around a rotating black hole has been Fourier transformed and integrated in
the frequency domain for a variety of situations where initial data played
no role (see Ref. \cite{FN89} for a review). Very recently it was
proved\cite{P97,CL97} that
nothing is intrinsically wrong with the Teukolsky equation when sources
extend to infinity and that a regularization method produces sensible results.
In order to incorporate initial
data and have a notable computational efficiency, concrete
progress has been made recently to complete a computational
framework that allows to
integrate the Teukolsky equation in the {\it time domain}: First, an
evolution code for integration of the Teukolsky wave equation is now
available\cite{KLPA97} and successfully tested\cite{CKL98}. Second, non
conformally flat Cauchy
data, compatible with Boyer-Lindquist slices of the Kerr geometry, began to
be studied with a Kerr-Schild\cite{BIMW97,MHS98} or an axially symmetric
\cite{BP98,KP98} ansatz. Finally, an
expression connecting $\psi_4$ to only Cauchy data, 
$$
\fbox{$\psi_4=\psi_4(h_{ij},K_{ij}),  
\partial_t\psi_4=\dot\psi_4(h_{ij},K_{ij})$}, 
$$
has been worked out
explicitly\cite{CL98I,CKL98,CLBKP98}. Note that those relashions hold
for the full spacetime and thus to any order in the perturbations
(see Lousto's proceeding contribution in this volume for a comprehensive
review on this problems).

Assuming that we can solve for the first order perturbations problem, we
decided to go one step forward in setting the formalism for the second order
perturbations. As motivations for this work we can cite the spectacular
results presented in Ref.\ \cite{GNPP96} for the head-on collision and the
hope to obtain similar agreement for the orbital binary black hole case
in the close limit. Second order perturbations of the Kerr metric
may even play a
more important role in this case since we expect the perturbative
parameter to be linear in the separation of the holes\cite{KP98b}
while in the head on case it is quadratic in the separation\cite{AP96}. The
nonrotating limit of our approach will also provide an independent test and
clarify some aspects of Ref.\ \cite{GNPP96} results. High precision
comparison with full numerical integration of Einstein equations using
perturbative theory as benchmarks is also one of the main
goals in this program.
However, the main mission of second order perturbations is to provide error
bars. It is well known that linearized perturbation theory does not provide,
in itself, any indication on how good the perturbative approximation is. In
fact, it is in general very difficult to estimate the errors involved in
replacing an exact solution of the full Einstein equations with an
approximate (perturbative) solution, i.e., to determine how small a
perturbative parameter $\varepsilon $ must be in order that the approximate
solution have sufficient accuracy. Moreover, first order perturbation theory
can be very sensitive to the choice of parametrization, i. e. different
choices of the perturbative parameter can affect the accuracy of the
linearized approximation \cite{AP96}. The only reliable procedure to resolve
the error and/or parameter arbitrariness is to carry out computations of the
radiated waveforms and energy to second order in the expansion parameter.
The ratio of second order corrections to the linear order results
constitutes the only direct and systematically independent measure of the
goodness of the perturbation results.

In Ref. \cite{CL99} we extend to second (and higher) order the Teukolsky
derivation of the equation that describes first order perturbations 
about a Kerr hole.
To do so we consider the Newman-Penrose \cite{NP62} formulation of the
Bianchi identities and Einstein equations for the Kinnersley null tetrad
(with $l^\mu$ and $n^\mu,m^\mu$ along the principal null directions), 
make a perturbative expansion of it, and decouple the equation that
describes the evolution of second (and higher) order perturbations.
This equation takes the following form 
\begin{equation}
\fbox{$\widehat{{\cal T}}\psi^{(2)}=S[\psi ^{(1)},\partial _t\psi ^{(1)}]$},
\label{uno}
\end{equation}
where $\psi\dot=(\rho^{(0)})^{-4}\psi_4$, 
$\widehat{{\cal T}}$ is the same (zeroth order) wave operator that
applies to first order perturbations \cite{T73} that in Boyer-Lindquist 
coordinates $(t,r,\vartheta ,\varphi )$, takes the following familiar 
form,
\begin{eqnarray}
&&\widehat{{\cal T}}=\left[ \frac{(r^2+a^2)^2}\Delta -a^2\sin ^2\vartheta
\right] \partial _{tt}+\frac{4Mar}\Delta \partial _{t\varphi }-4\left[
r+ia\cos \vartheta -\frac{M(r^2-a^2)}\Delta \right] \partial _t  \nonumber \\
&&-\,\Delta ^2\partial _r\left( \Delta ^{-1}\partial _r\right) -\frac 1{\sin
\vartheta }\partial _\vartheta \left( \sin \vartheta \partial _\vartheta
\right) -\left[ \frac 1{\sin ^2\vartheta }-\frac{a^2}\Delta \right] \partial
_{\varphi \varphi }  \label{master} \\
&&+\,4\left[ \frac{a(r-M)}\Delta +\frac{i\cos \vartheta }{\sin ^2\vartheta }%
\right] \partial _\varphi + \left( 4\cot ^2\vartheta +2\right) ,  \nonumber
\end{eqnarray}
where $M$ is the mass of the black hole, $a$ its angular momentum per unit
mass, $\Sigma \equiv r^2+a^2\cos ^2\vartheta $, and 
$\Delta \equiv r^2-2Mr+a^2$.
The source term  $S$ is quadratic in the first order perturbations:
\begin{eqnarray}
{\cal S} &=&2(\rho^{(0)})^{-4}\Sigma
\bigg\{\left[\left(\overline{\delta}+3\alpha +\overline{\beta}+4\pi-
\overline{\tau }\right)^{(0)}\left(\delta 
+4\beta -\tau\right)^{(1)}-\overline{d}_4^{(0)}\left( D+4\epsilon 
-\rho \right)^{(1)}\right]\psi _4^{(1)}  \nonumber \\
&+&\left[\left( \Delta +4\mu +\overline{\mu }+3\gamma 
-\overline{\gamma }\right)^{(0)}\left( \overline{\delta }+4\pi 
+2\alpha \right)^{(1)}\right] \psi _3^{(1)}\nonumber \\
&-&3\left[\psi _2^{(1)}
\Delta^{(0)}\lambda^{(1)}+\lambda^{(1)}\overline{d}_4^{(0)}
\psi _2^{(1)}\right]\bigg\},
\label{fuente}
\end{eqnarray}
where all the NP spin-coefficients and directional derivatives can be
expressed in terms of metric perturbations (see Ref.\cite{CL99} 
for details).

As we show explicitly in Ref. \cite{CL99}, $\psi_4^{(2)}$ is neither 
invariant under first order coordinates transformations nor 
second order tetrad rotations.
Thus, in order to integrate Eq.\ (\ref{uno}), one would have to evolve
$\psi^{(2)}$ in a fixed gauge (and tetrad) and then compute physical 
quantities, like radiated energy and waveform, in an asymptotically 
flat gauge.
This sort of approach was followed in Ref.\ \cite{GNPP98} to study
second order perturbations of a Schwarzschild black hole in the Regge-Wheeler
gauge which is a unique gauge that allows to invert expressions in terms
of generic perturbations and thus recover the gauge invariance.
There is not a generalization of the Regge-Wheeler gauge when studying
perturbations of a Kerr hole, essentially because one cannot perform a
simple multipole decomposition of the metric. Instead, Chrzanowski
\cite{C75} found two convenient gauges that allowed him to invert the
metric perturbations in terms of the Weyl scalars $\psi_4$ or $\psi_0$.
One can then use these metric perturbations to explicitly compute the 
source (\ref{fuente}), in terms of $\psi_4^{(1)}$ or $\psi_0^{(1)}$
only, which is the object we directly obtain from the integration of 
the first order Teukolsky equation.

The energy and momenta radiated at infinity to second perturbative order
can be computed using the standard methods of linearized gravity 
defined in asymptotically flat coordinates at future null infinity).
For outgoing waves the total radiated energy per unit time $(u=t-r)$
can thus be obtained from the Landau-Lifschitz pseudo tensor as 
\begin{equation}
\frac{dE}{du}=\lim_{r\to\infty}\left\{ \frac{r^2}{4\pi}%
\int_{\Omega}d\Omega\left| \int_{-\infty}^{u}d\tilde{u}\ \psi_4(\tilde{u%
},r,\vartheta,\varphi) \right|^2\right\}, \quad
d\Omega=\sin\vartheta\ d\vartheta\ d\varphi,
 \label{energy}
\end{equation}
where we can consider $\psi_4=\psi_4^{(1)}+\psi_{4}^{(2)\ AF}+...$

In the same way, one can also compute the total linear momentum radiated at
infinity per unit time along cartesian-like coordinates as\cite{NT80}
\begin{eqnarray}
&&\frac{dP_\mu}{du}=-\lim_{r\to\infty}\left\{ \frac{r^2}{4\pi}%
\int_{\Omega}d\Omega\ \tilde{l}_\mu\left| \int_{-\infty}^{u}d\tilde{u}\ 
\psi_4(\tilde{u},r,\vartheta,\varphi) \right|^2\right\},\label{momentum}\\
&&\tilde{l}_\mu=(1,-\sin\theta\cos\varphi,-\sin\theta\sin\varphi,-\cos\theta),
\nonumber
\end{eqnarray}
and the angular momentum carried away by the waves\cite{W80}
\begin{equation}
\frac{dJ_z}{du}=-\lim_{r\to\infty}\left\{ \frac{r^2}{4\pi}\ Re
\int_\Omega d\Omega
\left(\partial_\varphi\int_{-\infty}^{u}d\tilde{u}\
\psi_4(\tilde{u},r,\vartheta,\varphi) \right)
\left(\int_{-\infty}^{u}du^\prime\int_{-\infty}^{u^\prime}d\tilde{u}\ 
\overline{\psi}_4(\tilde{u},r,\vartheta,\varphi) \right)
\right\}.  \label{angmomentum}
\end{equation}

\bigskip

Higher than first order calculations are always characterized by an
extraordinary complexity and a number of subtle, potentially confusing,
gauge issues mainly due to the fact that a general second order gauge
invariant formulation is not yet at hand in the literature. In general,
gauge invariant quantities have an inherent physical meaning and they
automatically lead to the simpler and direct interpretation of the results.
In the Newman-Penrose formalism one has not only to look at gauge
invariance (i. e. invariance under infinitesimal coordinates
transformations), but also at invariance under tetrad rotations (see
Ref.\cite{CL99}). More specifically, the problem here is that the
waveform $\psi_4^{(2)}$ in Eq.\ (\ref{uno})) is neither first order
coordinate gauge invariant nor tetrad invariant. It is only invariant
under purely {\it second} order changes of coordinates, 
simply because $\psi_4$ vanishes on the background (Kerr metric). 
The question that arises therefore is whether $\psi _4^{(2)}$ can be 
unambiguously compared with, for instance, full numerical computations 
of the covariant $\psi_4^{Num}$.
Invariant objects to describe second perturbations lead us to reliable 
physical answers without having to face gauge difficulties.

To handle this problem we give an explicit and general prescription for the
construction of second order gauge and tetrad invariant objects representing
outgoing radiation and thus explictly build up a coordinate and tetrad
invariant quantity up to second order, $\psi_{I}^{(2)}$, which
has the property of reducing to the linear part
(in the second order perturbations of the metric)
of $\psi_4^{(2)}$ in an asymptotically flat gauge at the ``radiation
zone'', far from the sources. This property ensures us direct
comparison with $\psi_4^{Num}$ by constructing
$\psi_4^{(1)}+\psi_{4\ I}^{(2)}$. 
To do so we impose the waveform $\psi_{I}^{(2)}$
to be invariant under a ``combined'' transformation of both the
coordinates and the tetrad frame to first and second order.
The resulting second order invariant waveform can then be built
up out of the original $\psi_4^{(2)}$ plus corrections (quadratic in
the first order quantities) that cancel out the gauge and tetrad
dependence of $\psi_4^{(2)}$. The good of all this complex construction 
is that at the end $\psi_I$ fulfills a single wave equation of the 
same form as Eq.\ (\ref{uno})
\begin{equation}
\widehat{\cal T}[\psi^{(2)}+Q]=S+\widehat{\cal W}[Q]=
\fbox{$\widehat{\cal W}[\psi_I]=S_I$}, \label{final}
\end{equation}
being $\psi^{(2)}\dot=(\rho^{(0)})^{-4}\psi_4^{(2)}$ and 
$S_I$ a ``corrected'' source term build up out of (known to
this level) first order perturbations.

A number of interesting conceptual and technical issues raised from 
this computation, like the appearance of nonlocalities in the definition 
of the gauge invariant waveform when we want to relate it to known 
first order objects and its non uniqueness. Seen in retrospective, 
our method of generating
a gauge invariant object is like a machine that transforms any
(first order) gauge into an asymptotically flat
gauge, in particular, into the outgoing radiation gauge\cite{C75}. 
In fact,in this gauge we have 
$(\psi_4^{(2)})^{ORG}=(\psi_{4L}^{(2)})^{ORG}=
(\psi_I^{(2)})^{ORG}=\psi_I^{(2)}$.

The spirit of this work has been to show that there exists a gauge invariant
way to deal with second order perturbations in the more general case of a
rotating black hole and to provide theoretical support to the numerical
integration of the second order perturbation problem. In order to
implement such integration of Eq.\ (\ref{final}) (or equivalently of 
Eq.\ (\ref{uno})) we proceed as
follows: We assume that on an initial hypersurface we know the first and
second order perturbed metric and extrinsic curvature. We then solve the
first order problem, i.e. solve the standard Teukolsky equation for
$\psi_4^{(1)}$ (and for $\psi_0^{(1)}$).
Next we build up the perturbed metric coefficients in, for
instance, the outgoing radiation gauge (ORG). The perturbed
spin coefficients and the covariant basis are given by expressions in
terms of metric perturbations \cite{CL99}. 
Those are all the necessary elements to build
up the effective source term appearing on the right hand side of our
evolution equation.
For the computation of the radiated energy and momentum one uses
Eqs.\ (\ref{energy}) and (\ref{momentum}).
The advantage of this procedure is that we can now use the same
(2+1)-dimensional code for evolving the first order perturbations
\cite{KLPA97} by adding a source term.

An important application of the formalism presented here (see also 
Ref.\cite{CL99}) is to extend the numerical computations to the more
interesting case of rotating black holes in orbit around each other. 
The numerical integration of Eq.\ (\ref{final}) will be relevant not 
only for establishing the range of validity of the collision parameters
in the close limit approximation, but (hopefully) to produce a more
precise computation of the gravitational radiation. Direct
comparison with the existing codes for numerical integration of the
full nonlinear Einstein equations is possible.

%\section*{Acknowledgments}

\end{document}